\documentclass[letterpaper, 10 pt, conference]{ieeeconf}  

\IEEEoverridecommandlockouts                              

\usepackage{graphics} 
\usepackage{epsfig} 
\usepackage{mathptmx} 
\usepackage{times} 
\usepackage{amsmath} 
\usepackage{amssymb}  
\usepackage{algorithm} 
\usepackage{algpseudocode}
\usepackage{array}
\usepackage{epsfig}
\usepackage{amsmath}
\usepackage{graphicx}
\usepackage{tabto}
\usepackage{cleveref}
\usepackage{subfigure}
\usepackage{wrapfig}
\usepackage{textcomp}
\usepackage{float}
\usepackage{booktabs}
\usepackage{graphicx}
\usepackage{url}

\title{\LARGE \bf
Fast Approximate Time-Delay Estimation in Ultrasound Elastography
\\Using Principal Component Analysis 
}

\author{Abdelrahman Zayed$^{1}$ and Hassan Rivaz$^{2}$
\thanks{*This research was funded by Richard and Edith Strauss Foundation.}
\thanks{$^{1}$Abdelrahman Zayed is with Department of Electrical and Computer Engineering
and PERFORM Centre, Concordia University, Montreal, Quebec, Canada
        {\tt\small a\_zayed@encs.concordia.ca}}%
\thanks{$^{2}$Hassan Rivaz is with Department of Electrical and Computer Engineering
and PERFORM Centre, Concordia University, Montreal, Quebec, Canada
        {\tt\small hrivaz@ece.concordia.ca}}%
}

\begin{document}

\maketitle
\thispagestyle{empty}
\pagestyle{empty}

\begin{abstract}

Time delay estimation (TDE) is a critical and challenging step in all ultrasound elastography methods. A growing number of TDE techniques require an approximate but robust and fast method to initialize solving for TDE. Herein, we present a fast method for calculating an approximate TDE between two radio frequency (RF) frames of ultrasound. Although this approximate TDE can be useful for several algorithms, we focus on GLobal Ultrasound Elastography (GLUE), which currently relies on Dynamic Programming (DP) to provide this approximate TDE. We exploit Principal Component Analysis (PCA) to find the general modes of deformation in quasi-static elastography, and therefore call our method PCA-GLUE. PCA-GLUE is a data-driven approach that learns a set of TDE principal components from a training database in real experiments. In the test phase, TDE is approximated as a weighted sum of these principal components. Our algorithm robustly estimates the weights from sparse feature matches, then passes the resulting displacement field to GLUE as initial estimates to perform a more accurate displacement estimation. PCA-GLUE is more than ten times faster than DP in estimation of the initial displacement field and yields similar results.

\end{abstract}

\section{INTRODUCTION}

\label{sec:intro}
Ultrasound elastography is an emerging imaging modality that can estimate the mechanical properties of the tissue using Radio Frequency (RF) data. This is of great importance in the diagnosis of many diseases and has many clinical applications~\cite{gennisson2013ultrasound}. Elastography can be broadly classified into dynamic and quasi-static elastography~\cite{j2011recent}. Dynamic elastography often provides quantitative mechanical properties of the tissue. Quasi-static elastography, which is our concern in this paper, estimates the tissue deformation that occur due to applying a slowly varying external force \cite{parker2010imaging}.

Regardless of the elastography technique used, estimation of tissue displacement, also known as Time-Delay-Estimation (TDE), is a critical and challenging step. A growing interest is dedicated to developing robust and accurate TDE methods either using window-based~\cite{Jiang_2015,yuan2015analytical} or optimization-based~\cite{hashemi2017global,rivaz2011real} methods. Window-based techniques calculate the TDE for small windows of the RF data either by maximizing a similarity metric such as normalized cross correlation (NCC) or by finding the zero crossing of the phase of cross correlation \cite{yuan2015analytical}. Optimization-based techniques, such as Global Ultrasound Elastography (GLUE), calculate the TDE by minimizing a regularized cost function~\cite{hashemi2017global}.

An approximate estimation of the displacement map is mandatory in some methods (such as GLUE) or can help speed the search (such as in window-based methods). Dynamic Programming (DP)~\cite{rivaz2008ultrasound} has been often used to provide this initial estimate. However, GLUE requires initial estimates for all samples of the image and running DP for the entire image is computationally expensive. In this paper, we present a method that only needs DP initial estimates for few RF lines, and as such, is very computationally efficient.

Inspired by the success of~\cite{Wulff:CVPR:2015}, we first compute the principal components of deformation fields in quasi-static elastography by collecting many frames of ultrasound data during free-hand palpation from different phantoms and at different locations, in addition to \textit{in-vivo} data  collected from the liver. An ultrasound probe can move in 6 degrees of freedom (DOF), and we collect this data by allowing all kinds of motion. Although purely axial displacements of the probe are ideal for quasi-static elastography, we do not restrict our training database to this kind of motion to allow our model to better generalize to different users with different styles of compression. We then estimate any kind of compression as some weighted summation of these principal components. A recent work shows promising results of displacement estimation by learning a global dictionary of deformations in electrode displacement elastography~\cite{pohlman2018dictionary}.

\begin{figure}[b!]

\begin{minipage}[b]{1.0\linewidth}
\centering
\centerline{\includegraphics[width=0.3\linewidth,trim=0 0 0 0,clip]{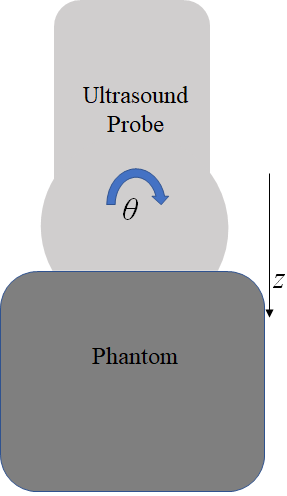}}
\centerline{(a) Directions of applied force}\medskip
\end{minipage}
\begin{minipage}[b]{1\linewidth}
\centering
\centerline{\includegraphics[width=1.1\linewidth,trim=0 0 0 0,clip]{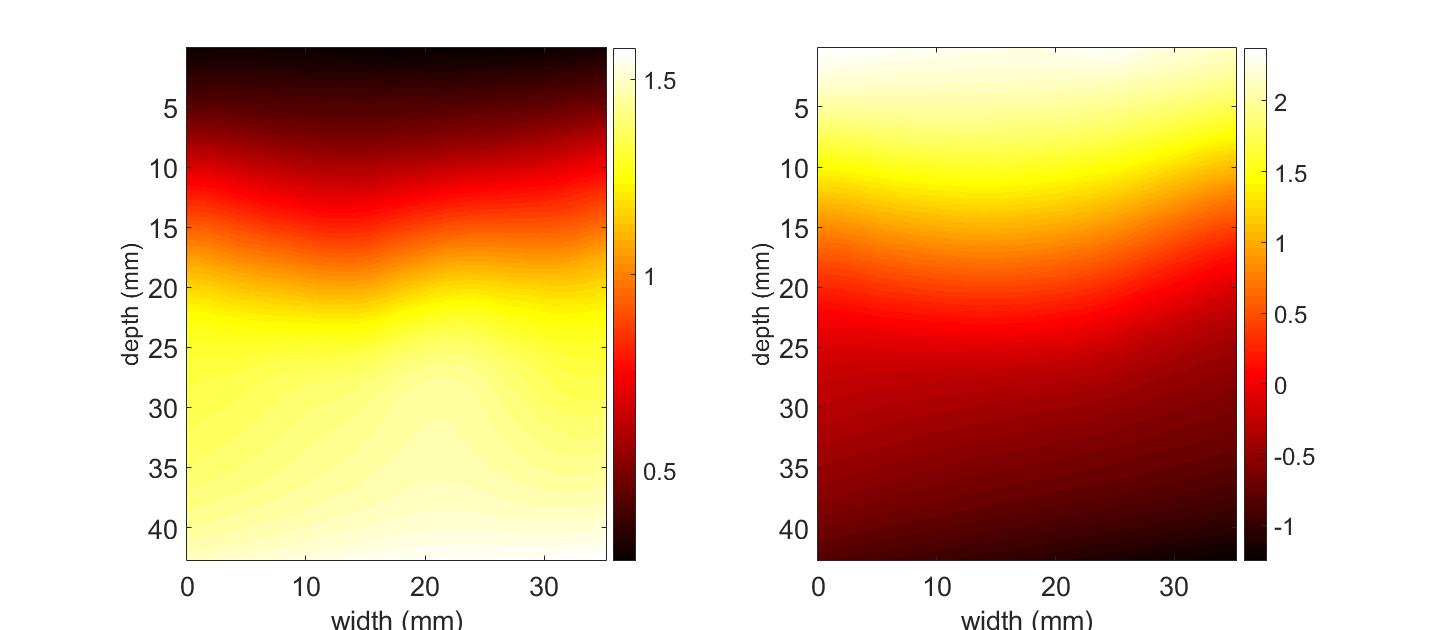}}
\centerline{(b) Axial deformation ($z$)}\medskip
\end{minipage}
\begin{minipage}[b]{1\linewidth}
\centering
\centerline{\includegraphics[width=1.1\linewidth,trim=0 0 0 0,clip]{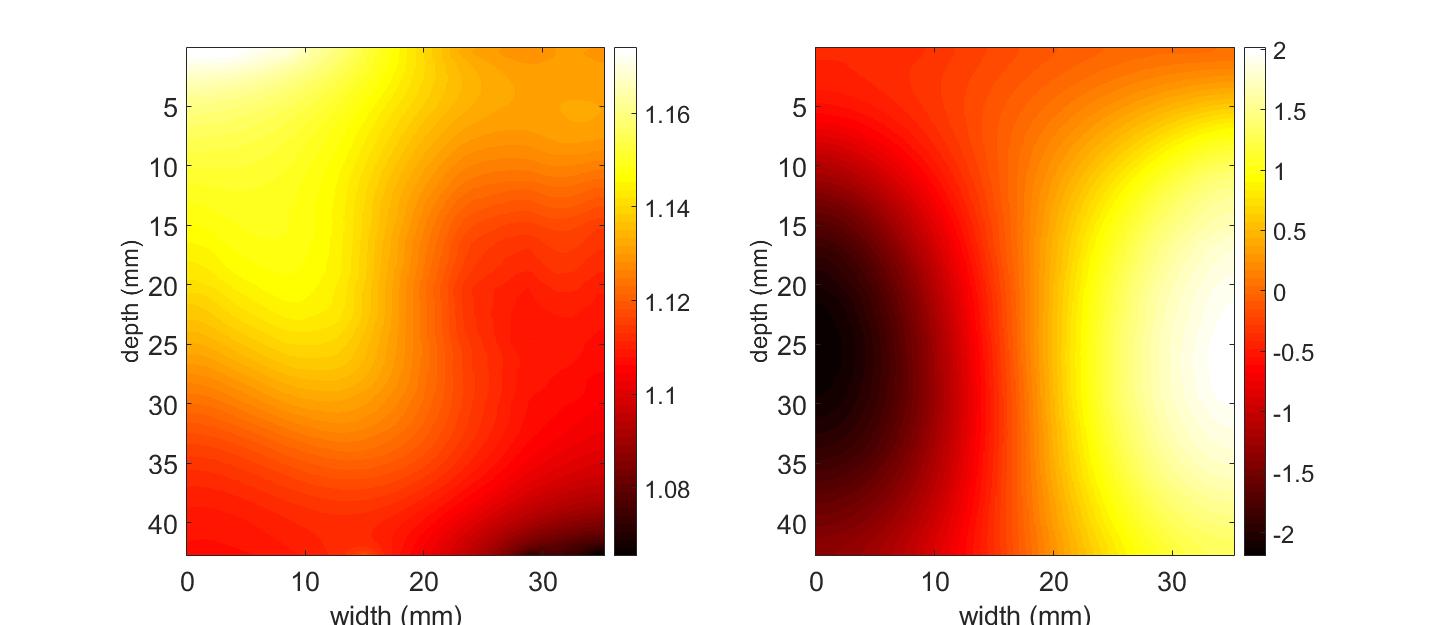}}
\centerline{(c) In-plane rotation ($\theta$)}\medskip
\end{minipage}
\caption{Principal components of in-plane axial displacement learned from both \textit{in-vivo} and phantom experiments. In (a), translation of the probe along $z$ and its rotation by $\theta$ generates axial deformation in the phantom. In (b), extension and compression principal components along $z$ are shown. In (c), displacement arising from rotation by $\theta$ is shown.}
\label{fig0}
\end{figure}

\section{METHOD}

DP is a global optimization method that can provide discrete displacement estimates~\cite{rivaz2008ultrasound}. Although substantially more efficient than a brute force algorithm, DP is computationally expensive. Herein, we propose a data-driven technique that learns the deformation modes (i.e. principal components) of quasi-static elastography, and then approximates any deformation as a weighted sum of these modes. 

During training, we calculate $N$ principal components that describe the axial displacement as the tissue deforms. These principal components are represented by $\textbf{b}_1$ to $\textbf{b}_{N}$. Fig. \ref{fig0} shows a schematic of probe motion and some of these principal components learned from real experiments. It is clear that applying an axial force produces an axial displacement that increases linearly with the depth, starting from zero at the probe level, while the opposite happens after the removal of the force. As for rotation, it produces an axial displacement in two opposite directions (upwards and downwards) on both sides.

Let $I_1$ and $I_2$ be RF frames of size $m\times l$, where $m$ is the number of samples in an RF line and $l$ is the number of RF lines. Our goal is to calculate an approximate deformation field between these two frames.
Algorithm~\ref{algorithm:pca_glue} shows the steps of our method, which is also outlined below.
\subsection{Calculating the displacement of the sparse features}

We first choose $p < < l$ equidistant RF lines and run DP to calculate an integer axial displacement field. DP produces $K = m \times p$ displacements between $I_1$ and $I_2$. Compared to the image size $m \times l$, $K$ represents a sparse set of correspondences.
We perform a simple linear interpolation to convert these integer deformation estimates to real numbers. We then form vector \textbf{c} of length $K$ that corresponds to the interpolated axial displacements.

\subsection{Estimating the displacement in the whole image using the principal components}

We form the matrix \textbf{A} such that
\begin{equation}
\textbf{A}=
\begin{bmatrix}
\textbf{ b}_1(q_1) &\textbf{ b}_2(q_1) & \textbf{ b}_3(q_1) & \dots &\textbf{ b}_N(q_1) \\
\textbf{ b}_1(q_2) & \textbf{ b}_2(q_2) & \textbf{ b}_3(q_2) & \dots & \textbf{ b}_N(q_2) \\
\hdotsfor{5} \\
\textbf{ b}_1(q_K) & \textbf{ b}_2(q_K) & \textbf{ b}_3(q_K) & \dots & \textbf{ b}_N(q_K)
\end{bmatrix}
\end{equation}
where $q_1$ to $q_K$ correspond to our 2D sparse features chosen along the $p$ RF lines before deformation. $N$ is the number of principal components used. 
We then solve the optimization equation below:
\begin{equation}
\label{lsq}
\hat{\textbf{w}}= \textrm{arg}\min_\textbf{w} ||\textbf{Aw--c}|| 
\end{equation}
The intuition behind this equation is that we find the weight vector $\textbf{w}=(w_1,...,w_N)^T$ that best describes the displacement of the $p$ RF lines as a linear combination of the principal components that we computed offline. Hence, the displacement of the rest of the RF lines is assumed to be also a linear combination of the same principal components weighted by the same weight vector.

Consequently, the initial estimate of the axial displacement \textbf{\^{d}} is calculated as:
\begin{equation}
\label{eq2}
\textbf{\^{d}} = \sum_{n=1}^{N} \hat{w_n} \textbf{b}_n
\end{equation}
where \textbf{\^{d}} is passed to GLUE to calculate the fine-tuned \textbf{d}.
Finally, spatial differentiation is performed on \textbf{d} to obtain the strain image.

\begin{algorithm}[!t]
	\caption{PCA-GLUE. Disp. refers to displacement.}
	\begin{algorithmic}[1]
		\Procedure {PCA-GLUE~}{}
		\State \text{Choose  $p$ equidistant RF lines}
		\State \text{Run DP to get the integer axial disp. of the $p$ RF lines}
		\State \text{Solve Eq.~\ref{lsq} to get the vector $\textbf{w}$}
		\State \text{Compute the initial axial disp. of all RF lines by Eq.~\ref{eq2}}
		\State \text{Use GLUE to calculate the exact axial disp.}
		\State \text{Strain is obtained by spatial differentiation of the disp.}
		\EndProcedure
	\end{algorithmic}
	\label{algorithm:pca_glue}
	\vspace{.12cm}
	\vspace{3mm}
\end{algorithm}

\subsection{Data Collection and Principal Components }
\subsubsection{Training the Model}
For our training data, we collected approximately 4000 frames from both \textit{in-vivo} liver data and 3 different CIRS phantoms (Norfolk, VA), namely Models 040GSE,  039 and  059 at different locations. RF phantom data were collected using a 12R Alpinion Ultrasound machine (Bothell, WA) with an L3-12H high density linear array probe at a center frequency of 8.5 MHz and sampling frequency of 40 MHz. The probe can move in 6 degrees of freedom (DOF) in total, wherein 3 result in out-of-plane motion and 3 lead to in-plane motion. Hence, we performed 3 types of probe motion that lead to in-plane motion, namely purely axial, purely in-plane rotation and purely lateral displacement (two of these DOFs are shown in Fig. \ref{fig0}). We used GLUE \cite{hashemi2017global} to calculate the displacement images for training. It took around 5 hours to extract the principal components, but this step is performed only once offline and does not need to be performed at test time.

\subsubsection{Principal Components}
More principal components better capture the space of different deformations. However, this will render the algorithm more computationally expensive and also less robust to noise. To specify the number of principal components, we choose $N$ principal components and calculate the variance in our data after being projected to a lower dimensional
space. In our case, taking just $N = 12$ principal components was enough to preserve 95\% of the variance in the high-dimensional data.

As for the lateral displacement, we tried to calculate their own principal components, but that was not possible. The variance was almost equally distributed over all the eigenvalues (resembling  white noise). Therefore, to preserve up to 95\% of the variance, we had to take hundreds of principal components. That is because GLUE is not very accurate in calculating  lateral displacements, mainly because of the poor resolution of ultrasound in this direction. As a result, instead of calculating principal components in the lateral direction, we simply perform a bi-linear interpolation of the $K=m\times p$ lateral displacement estimates. Our results show that this bi-linear interpolation provides acceptable lateral displacement estimates.

\section{EXPERIMENTS AND RESULTS}
We chose $N$=12 and $p$=5 for the number of principal components and RF lines respectively, which corresponds to the best trade-off between running time and accuracy. In order to quantitatively measure the performance of our method, we used two unitless metrics which are signal to noise ratio (SNR) and contrast to noise ratio (CNR)~\cite{ophir1999elastography}:
\begin{equation}
CNR=\frac{C}{N}=\sqrt{\frac{2(\bar{s_{b}}-\bar{s_{t}})^2}{\sigma_{b}^2 + \sigma_{t}^2} }, SNR=\frac{\bar{s}}{\sigma}
\end{equation}\\
where $ \bar{s_{t}}$ and $ \sigma_{t}^2$ are the strain average and variance of the target window, $ \bar{s_{b}}$ and $ \sigma_{b}^2$ are the strain average and variance of the background window respectively. To calculate the SNR, we used the same background window, which means that $\bar{s}$=$ \bar{s_{b}}$ and $\sigma $=$ \sigma_{b}$.

Our unoptimized implementation of the proposed method in MATLAB takes approximately 258 ms on an 8th generation 3.2 GHz Intel core i7 to estimate the 2D displacement fields between two very large images of size $2304 \times 384$. In comparison, DP takes approximately 2.6 seconds on the same system to estimate the 2D displacement fields of two images of the same size, more than ten times longer than the proposed method. Further speed-up can be achieved by optimizing our code and implementing it in C.

\begin{figure}[h]
\begin{center}
\subfigure[PCA-GLUE]{\includegraphics[width=.44\linewidth]{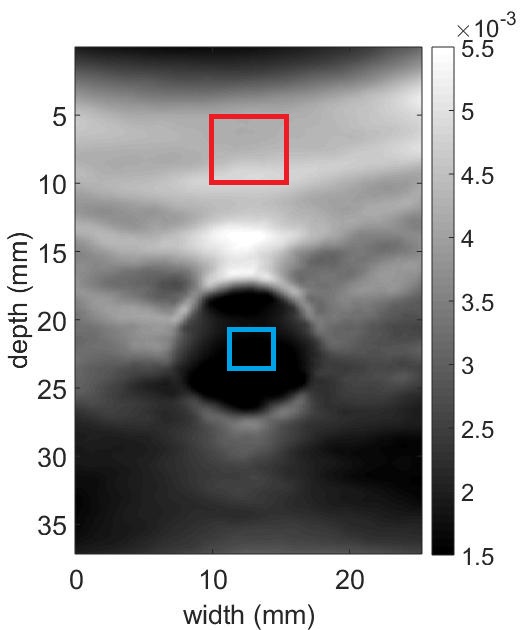}}
\subfigure[GLUE]{\includegraphics[width=.44\linewidth]{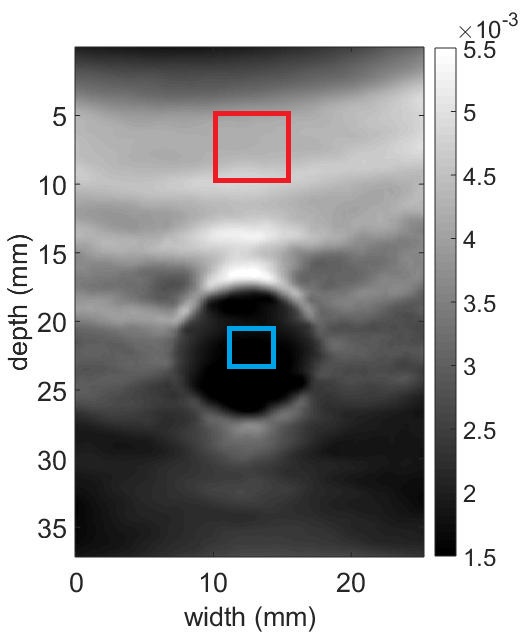}}
\end{center}%
\caption{Results of the axial strain images for the real phantom experiment. The target and background windows are used for  calculating SNR and CNR.}%
\vspace{-.1cm}
\label{fig1}%
\end{figure}

\begin{figure}[h]
\begin{center}
\subfigure[PCA-GLUE]{\includegraphics[height=4.5 cm,width=4.26 cm]{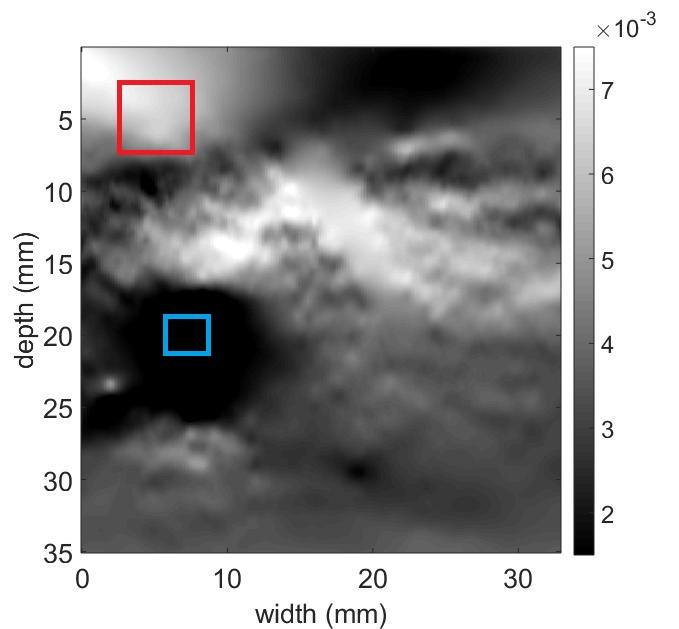}}
\subfigure[GLUE]{\includegraphics[height=4.5 cm,width=4.26 cm]{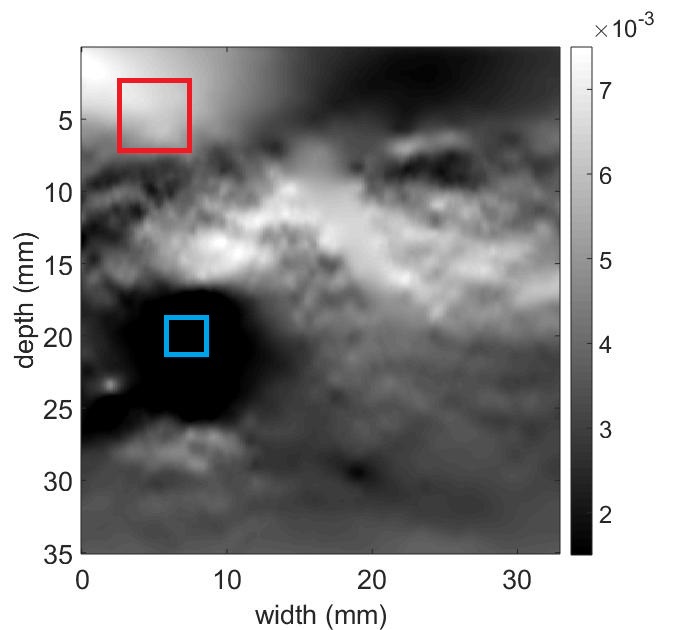}}
\end{center}%
\caption{Results of the axial strain images for \textit{in-vivo} data. The target and background windows are used for  calculating SNR and CNR.}%
\vspace{-.1cm}
\label{fig11}%
\end{figure}

\begin{figure*}[h]
\begin{center}

\subfigure[PCA-GLUE using 5 RF lines]{\includegraphics[width=.22\linewidth,trim=8 0 5 0,clip]{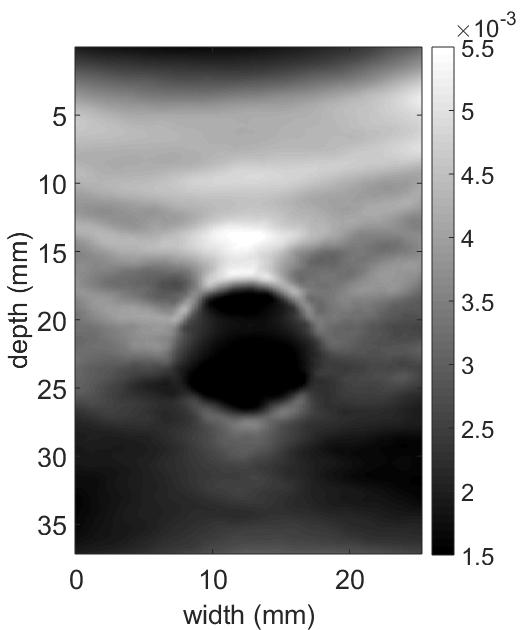}}\quad
\subfigure[PCA-GLUE using 15 RF lines]{\includegraphics[width=.22\linewidth,trim=8 0 5 0,clip]{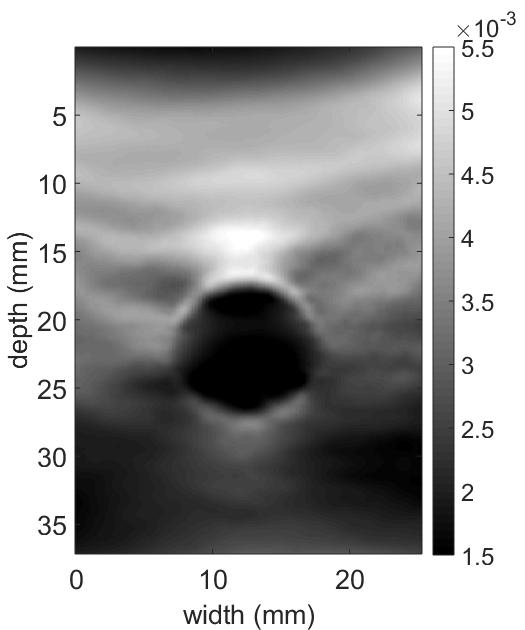}}\quad
\subfigure[PCA-GLUE using 30 RF lines]{\includegraphics[width=.22\linewidth,trim=8 0 5 0,clip]{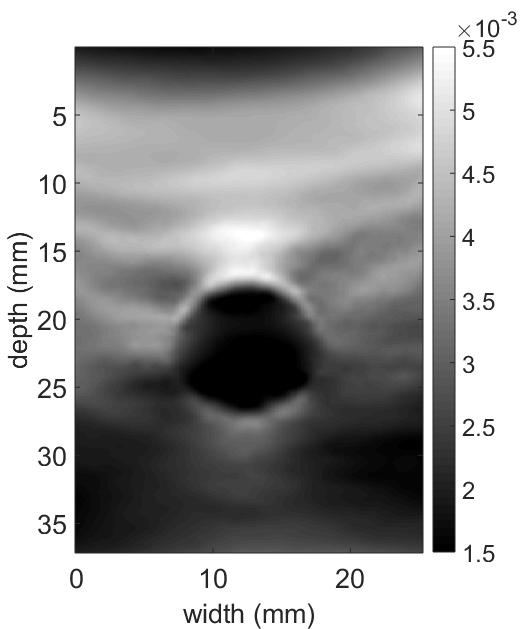}}\quad
\subfigure[GLUE]{\includegraphics[width=.22\linewidth,trim=8 0 5 0,clip]{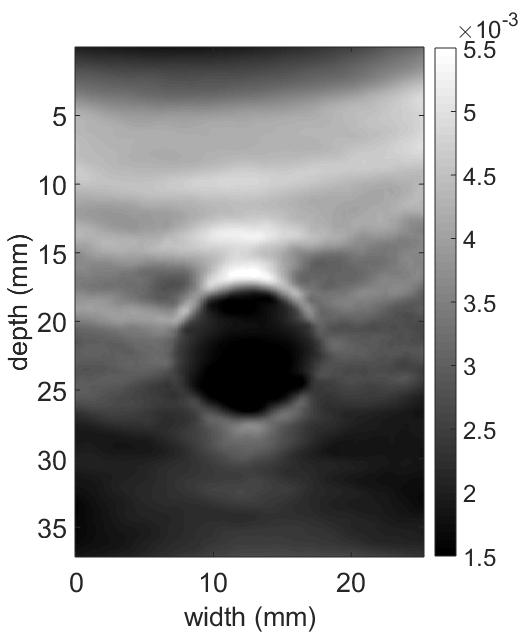}}\quad
\end{center}%
\caption{Results of the axial strain images for the phantom experiment as we increase the number of RF lines $p$ from 5 to 30.}%
\label{fig2}%
\end{figure*}
\subsection{Phantom Results}
For our experiments, we used data acquired from the same three CIRS elastography phantoms using the 12R Alpinion E-Cube  ultrasound machine at a sampling rate of 40 MHz. Fig. \ref{fig1} shows the results of both GLUE and PCA-GLUE, as well as the windows used as target and background. Table \ref{tab1} summarizes the results, where the values of SNR and CNR for both  GLUE and PCA-GLUE are almost the same. It is important to note that the reported results are on new test data which our model has not seen before during the training phase.

\subsection{\textit{In-vivo} Results}
Our \textit{in-vivo} results were collected from one patient undergoing open surgical radiofrequency thermal ablation for primary or secondary liver cancers. The data  was acquired from Johns Hopkins Hospital. Fig. \ref{fig11} shows the results of both GLUE and PCA-GLUE, as well as the windows used as target and background. Table \ref{tab2} summarizes the results, where the values of SNR and CNR for both  GLUE and PCA-GLUE are very close.

\subsection{Varying the number of sparse features $K$}
Increasing $p$ means that there are more correspondences available for Eq.~\ref{lsq}, leading to more accurate weights. However, DP should be run on more RF lines which increases the computational complexity. Fig. \ref{fig2} shows that increasing the number of RF lines above 5 will not change the results, and therefore, we set $p = 5$ RF lines.

\begin{table}
\centering
\caption{The SNR and CNR values of the axial strain images for the phantom experiment. Target windows and background windows are of size $3$ $mm$ $\times$ $3$ $ mm$ and $5$ $mm$ $\times$ $5$ $ mm$ respectively as shown in Fig. \ref{fig1}. SNR is calculated for the background window.}
\begin{tabular}{lrr}
\toprule
\textbf{Method used} & \textbf{SNR} & \textbf{CNR} \\
\midrule
GLUE & 22.51 & 20.74 \\
PCA-GLUE & 22.50 & 20.75 \\
\bottomrule
\end{tabular}
\label{tab1}
\end{table}

\begin{table}
\centering
\caption{The SNR and CNR values of the axial strain images for the \textit{in-vivo} data. Target windows and background windows are of size $3$ $mm$ $\times$ $3$ $ mm$ and $5$ $mm$ $\times$ $5$ $ mm$ respectively as shown in Fig. \ref{fig11}. SNR is calculated for the background window.}
\begin{tabular}{lrr}
\toprule
\textbf{Method used} & \textbf{SNR} & \textbf{CNR} \\
\midrule
GLUE & 20.26 & 17.18 \\
PCA-GLUE & 20.49 & 18.09 \\
\bottomrule
\end{tabular}
\label{tab2}
\end{table}

\section{DISCUSSION AND CONCLUSIONS}

PCA-GLUE provides an initial displacement map substantially faster than DP. A second important application of our proposed framework is that the weights of the principal components determine the magnitude of different types of probe motion (e.g. purely axial, in-plane rotation, etc.). For quasi-static elastography, purely axial displacements usually lead to  strain images of higher quality. Therefore, our proposed method can be used as a very fast technique for selecting good frames for further processing.

In this paper, we introduced a novel data-driven method for efficiently calculating an initial axial displacement between two frames of RF data.
In quasi-static elastography, there is a large variability in the types of probe motion as well as in tissue inhomogeneities. This leads to a large variability in the underlying displacement field. However, this displacement field can be approximated with a weighted summation of few principal components of tissue deformation that capture most of displacement variability.  Our proposed method is more than 10 times faster than DP and generates similar displacement estimates.

\addtolength{\textheight}{-12cm}   




\section*{ACKNOWLEDGMENT}

The \textit{in-vivo} data were collected at Johns Hopkins Hospital. The authors would like to thank the principal investigators Drs. E. Boctor, M. Choti and G. Hager who provided us with the data.

\bibliography{refs}
\end{document}